\documentclass[11pt]{article}
\usepackage{comment,soul}
\usepackage[hidelinks]{hyperref}
\usepackage{enumitem}
\usepackage{amsmath,amssymb,epsf,cite,graphicx,subfigure}
\usepackage{amsmath,braket,tikzsymbols}
\usepackage[bbgreekl]{mathbbol}
\usepackage{comment}
\usepackage[export]{adjustbox}
\usepackage{dsfont}
\usepackage{faktor}
\usepackage{mathrsfs}
\usepackage{float}
\usepackage{empheq}

\numberwithin{equation}{section}

\definecolor{airforceblue}{rgb}{0.36, 0.54, 0.66}
\newcommand{\beq}{\begin{equation}}
\newcommand{\eeq}{\end{equation}}

\newcommand{\tr}{\text{tr}}
\newcommand{\vol}{\text{vol}}
\newcommand{\dd}{\text{d}}

\begin{document}
\baselineskip=15.5pt
\pagestyle{plain}
\setcounter{page}{1}

\begin{center}
{\LARGE \bf Higher-loop norm of the no-boundary state}
\vskip 1cm

\textbf{Jordan Cotler}

\vspace{0.5cm}

{Department of Physics, Harvard University, Cambridge, MA 02138, USA \\}

\vspace{0.3cm}

{\tt  jcotler@fas.harvard.edu}

\medskip

\end{center}

\vskip1cm

\begin{center}
{\bf Abstract}
\end{center}
\hspace{.3cm}

The leading contribution to the de Sitter no-boundary state comes from geometries with spherical spatial slices, including the Hartle-Hawking geometry and fluctuations around it. Recent work showed that this leading contribution has vanishing norm at one loop.  Here we show that the norm in fact vanishes to all orders in perturbation theory.

\newpage

\tableofcontents

\section{Introduction}

The no-boundary state is an appealing candidate for the initial quantum state of the universe~\cite{Hartle:1983ai}.  From the perspective of the gravitational path integral, the no-boundary proposal defines a wavefunction by summing over smooth, compact geometries that realize specified boundary data and otherwise cap off regularly in the interior. In this work we consider $\Lambda > 0$ Einstein gravity and focus on the late-time no-boundary state, defined by specifying boundary data at future infinity and summing over geometries that approach de Sitter asymptotics there.  The leading, least-action contribution is furnished by spacetimes with spherical spatial slices, including the Hartle-Hawking saddle. It is therefore natural to focus on this leading contribution, since it is taken to dominate the semiclassical approximation to the no-boundary state. A substantial literature has analyzed this saddle when coupled to matter, including inflatons, with the future asymptotically de Sitter region interpreted as the endpoint of an inflationary de Sitter phase (see e.g.~\cite{Halliwell:1984eu, Hartle:2008ng, Hartle:2010vi, Lehners:2023yrj} as well as~\cite{Maldacena:2024uhs} for a recent discussion). Although there have been numerous phenomenological discussions of the no-boundary proposal, some basic quantum properties of the leading contribution have only recently been computed~\cite{Chakraborty:2023yed, Chakraborty:2023los, Cotler:2025gui}.

In recent work~\cite{Cotler:2025gui}, the norm of the leading contribution to the no-boundary state was computed in pure Einstein gravity to one loop in gravitational fluctuations and was found to vanish in $d \geq 3$ spacetime dimensions. In $d = 3$ spacetime dimensions, the norm was further shown to vanish to all loop orders, although that argument does not directly generalize to higher dimensions. These conclusions are unchanged upon coupling gravity to matter placed jointly in the no-boundary state. This leaves open the question of whether the norm of the leading contribution to the no-boundary state vanishes to \emph{all loops} in general dimension $d \geq 3$. In this paper we show that it does.

Let us first recall the result of~\cite{Cotler:2025gui}.  In $d$ spacetime dimensions, the leading contribution to the no-boundary state can be written schematically as
\begin{align}
\Psi[\gamma] = \int_{\text{no}\,\text{bdy}}^\gamma  \frac{[\dd g]}{\text{diffs}}\,e^{i S_{\text{EH}}[g]},
\end{align}
where $\gamma$ is a metric on the future boundary $\mathbb{S}^{d-1}$ defined modulo $\text{diff}\times\text{Weyl}$, and $S_{\text{EH}}$ is the Einstein-Hilbert action plus boundary terms.  The inner product of $\Psi[\gamma]$ with $\Psi^*[\gamma]$ is then given by
\begin{align}
\label{E:IP1}
\int \frac{[\dd\gamma]}{\text{diff}\times\text{Weyl}} \,|\Psi[\gamma]|^2\,,
\end{align}
which has as a saddle point $\gamma = \gamma_0$\,, namely the round metric on $\mathbb{S}^{d-1}$, where the corresponding bulk geometry is the Hartle-Hawking geometry.  To proceed, we can expand around the saddle as $\gamma = \gamma_0 + \sqrt{G}\, \mathfrak{h}$, and compute~\eqref{E:IP1} perturbatively in fluctuations.  In order to fix $\text{diff}\times\text{Weyl}$, we can impose a transverse-traceless gauge $\nabla_{\!\gamma_0}^j \mathfrak{h}_{ij} = 0$ and $\gamma_0^{ij} \mathfrak{h}_{ij} = 0$ using the Faddeev-Popov procedure.  At one loop, the associated ghosts have zero modes corresponding to residual unfixed gauge transformations, which here are conformal Killing vectors (CKVs) that generate an $SO(d,1)$ symmetry.  To account for these residual gauge transformations we should perform an additional gauge fixing.  Equivalently, since $\Psi_{\text{1-loop}}[\gamma_0 + \sqrt{G} \,\mathfrak{h}]$ is $SO(d,1)$-invariant and has no bosonic zero modes to quadratic order in $\mathfrak{h}$-fluctuations, we can simply divide by the volume of the residual gauge transformations `by hand'.  Therefore the norm at one loop is proportional to $1/\vol(SO(d,1))$, which is zero since $SO(d,1)$ is non-compact~\cite{Cotler:2025gui}.  The appearance of this volume factor is familiar from the one-loop computation of the gravity $\mathbb{S}^d$ partition function, but there one finds $1/\vol(SO(d+1))$~\cite{Anninos:2020hfj}, which is finite.\footnote{The $1/\vol(SO(d,1))$ in the inner product analysis is in part a consequence of imposing that $\gamma$ is a \emph{real} metric and that we are dividing by \emph{real} $\text{diff}\times\text{Weyl}$ transformations.  If we analytically continued $SO(d,1)$ to $SO(d+1)$ then the analytically continued boosts would act as \emph{complex} $\text{diff}\times\text{Weyl}$ transformations on $\gamma$.}  It has been subsequently noted that the vanishing of the norm can be consistently reinterpreted in the group-averaging approach~\cite{Held:2025mai}. 

In $d = 3$, the only metric on $\mathbb{S}^2$ modulo $\text{diff}\times\text{Weyl}$ is the round one $\gamma_0$, and so one can show that the norm is zero to all loops in that setting.  For $d > 3$ the story is more complicated.  In transverse-traceless gauge, the CKVs corresponding to boosts fail to be ghost zero modes at higher loops; however, this does not rule out that there are more complicated residual gauge transformations that together furnish a representation of $\mathfrak{so}(d,1)$.

In the remainder of this paper, we successfully find the perturbative nonlinear completion of the ghost zero modes in transverse-traceless gauge. Specifically, we find a family of \emph{field-dependent} $\text{diff} \times \text{Weyl}$ transformations $\delta_A$ which, to all orders in perturbation theory, are both ghost zero modes and provide a representation of $\mathfrak{so}(d,1)$.  That is, $[\delta_A, \delta_B] = f_{\!AB}^{\,\,\,\,\,\,\,\,C}\, \delta_C$, where $f_{\!AB}^{\,\,\,\,\,\,\,\,C}$ are the $\mathfrak{so}(d,1)$ structure constants, and moreover at leading order each $\delta_A$ is just the $\text{diff} \times \text{Weyl}$ transformation generated by an ordinary conformal Killing vector, together with its associated Weyl rescaling.  Furthermore the all-loop wavefunction is invariant under the $\delta_A$'s and has no bosonic zero modes, and hence the norm is proportional to $1/\vol(SO(d,1))$ at all loops.

We will conclude our paper with a discussion of possible physical implications of our findings.  In the Appendix we give a simple argument that the gravity $\mathbb{S}^{d}$ partition function is proportional to $1/\vol(SO(d+1))$ at all orders in perturbation theory, and also provide an analogous argument for Yang-Mills theories.  These arguments are simple because Weyl transformations do not enter, and the residual gauge redundancy is generated by the ordinary Killing vectors (in gravity) or constant gauge transformations (in Yang-Mills) to all orders in perturbation theory.

\section{Higher-loop norm}

In this Section, we consider the all-loop norm using the transverse-traceless gauge fixing, and show that the all-loop norm of the leading contribution to the no-boundary state vanishes.  Then we discuss why the vanishing of the norm persists upon coupling to matter.

\subsection{Transverse-traceless gauge fixing}
\label{subsec:TT}

Consider metrics on $\mathbb{S}^{d-1}$ which we parameterize by $\gamma_{ij} = \gamma_{0ij} + \sqrt{G}\,\mathfrak{h}_{ij}$.  We write the transverse-traceless gauge-fixing conditions as
\begin{align}
\label{E:FHeqs1}
F_i[\gamma] \equiv \nabla_{\!\gamma_0}^j \mathfrak{h}_{ij} = 0\,,\qquad H[\gamma] = \gamma_0^{ij} \mathfrak{h}_{ij} = 0\,,
\end{align}
where the $\nabla_{\!\gamma_0}^j$ is raised by $\gamma_0$.  These constraints carve out the gauge-fixing slice
\begin{align}
\mathcal{S} \equiv \{\gamma\,:\, F_i[\gamma]=0,\, H[\gamma]=0\}\,.
\end{align}
The variations of $F_i$ and $H$ under a boundary diffeomorphism $\xi$ and Weyl transformation $\sigma$ are
\begin{align}
\label{E:deltaF0}
\delta_{\xi, \sigma} F_i &= \nabla_{\!\gamma_0}^j\!\left(\nabla_{\!\gamma_0\,i} \xi_j + \nabla_{\!\gamma_0\,j} \xi_i + 2 \gamma_{0ij} \sigma\right) + O(\sqrt{G}) \\
\label{E:deltaH1}
\delta_{\xi, \sigma} H &= 2\left(\nabla_{\!\gamma_0}^i \xi_i + (d-1)\sigma\right) + O(\sqrt{G})\,.
\end{align}
In the Faddeev-Popov construction, the ghost action is obtained by linearizing the gauge-fixing conditions along the $\text{diff}\times\text{Weyl}$ orbit.  At one loop this reduces to the kernel of the linearized Faddeev-Popov operator around $\mathfrak{h} = 0$, with zero modes corresponding to the usual conformal Killing vectors on $\mathbb{S}^{d-1}$.  We now recall this computation to develop notation that we will build upon.

Let us compute the zero modes of the leading order variations of the gauge-fixing conditions in~\eqref{E:deltaF0} and~\eqref{E:deltaH1}.  Setting $\delta_{\xi,\sigma}H\big|_{F_i=H=0} = 0$, at leading order we find
\begin{align}
\sigma = - \frac{1}{d-1} \nabla_{\!\gamma_0}^i \xi_i\,,
\end{align}
and so plugging this in to $\delta_{\xi,\sigma}F_i\big|_{F_i=H=0} = 0$ gives, at leading order,
\begin{align}
\label{E:CKVeq0}
-\nabla_{\!\gamma_0}^j\!\left(\nabla_{\!\gamma_0\,i} \xi_j + \nabla_{\!\gamma_0\,j} \xi_i - \frac{2}{d-1} \gamma_{0ij} \nabla_{\gamma_0}^k \xi_k\right) = 0\,.
\end{align}
It is prudent to define the conformal Killing operator $\textsf{C}$, which acts on 1-forms by
\begin{align}
(\textsf{C}\xi)_{ij} \equiv \nabla_{\!\gamma_0\,i} \xi_j + \nabla_{\!\gamma_0\,j} \xi_i - \frac{2}{d-1} \gamma_{0ij} \nabla_{\gamma_0}^k \xi_k\,.
\end{align} 
The operator $\textsf{C}$ is evidently a map $\textsf{C} : \Omega^1 \to \text{ST}_2$ from 1-forms to symmetric traceless 2-tensors, and famously $\textsf{C} \xi = 0$ if and only if $\xi$ is a CKV.  We will also utilize the adjoint of $\textsf{C}$, which is constructed as follows.  For $\xi_i$ a 1-form and $S_{ij}$ a symmetric, traceless tensor, we can form the inner product
\begin{align}
\langle S_{ij}, (\textsf{C} \xi)_{ij}\rangle_{\text{ST}_2} &\equiv \frac{1}{2}\int \dd^{d-1} x \sqrt{\gamma_0}\,S^{ij} (\textsf{C} \xi)_{ij} = - \int \dd^{d-1} x \sqrt{\gamma_0}\,(\nabla_{\!\gamma_0\,j} S^{ij})\,\xi_i \equiv \langle (\textsf{C}^\dagger S)_i, \xi_i\rangle_{\Omega^1}\,,
\end{align}
where in going to the second equality we have integrated by parts and used that $S_{ij}$ is symmetric and traceless.  On the far right-hand side, we have
\begin{align}
(\textsf{C}^\dagger S)_{i} \equiv - \nabla_{\!\gamma_0}^j S_{ij}\,,
\end{align}
where $\textsf{C}^\dagger : \text{ST}_2 \to \Omega^1$.  Defining $\textsf{D} := \textsf{C}^\dagger \textsf{C}$, we have that $\textsf{D}$ is positive semidefinite and has zero modes corresponding to the CKVs.  In fact, the zero modes of $\textsf{D}$ are \emph{exactly} the CKVs.\footnote{Explicitly, $\langle\xi,\textsf{D}\xi\rangle_{\Omega^1}=\langle \textsf{C}\xi,\textsf{C}\xi\rangle_{\text{ST}_2}\ge 0$ implies $\textsf{D}\xi=0$ if and only if $\textsf{C}\xi=0$, so $\ker(\textsf{D})=\ker(\textsf{C})$ and therefore the only zero modes of $\textsf{D}$ are the CKVs.}  Then~\eqref{E:CKVeq0} is equivalent to
\begin{align}
(\textsf{D} \xi)_i = 0\,,
\end{align}
implying that $\xi_i$ is a CKV, and so the one-loop ghost zero modes correspond precisely to the CKVs $K_A$ together with their associated Weyl rescalings.

Our analysis above recapitulates the result of~\cite{Cotler:2025gui} that at one loop, the ghost sector is invariant under precisely the $\text{diff}\times\text{Weyl}$ transformations $\delta_{K_A, \sigma_A}$ with $\sigma_A = - \frac{1}{d-1}(\nabla_{\gamma_0} \cdot K_A)$, which satisfy
\begin{align}
\label{E:comm1}
\big[\delta_{K_A, \,\sigma_A}\,,\, \delta_{K_B, \,\sigma_B}\big] = f_{\!AB}^{\,\,\,\,\,\,\,\,C} \,\delta_{K_C, \,\sigma_C}\,,
\end{align}
where $f_{\!AB}^{\,\,\,\,\,\,\,\,C}$ are the $\mathfrak{so}(d,1)$ structure constants.  The commutator follows from the usual identity $[\delta_{\xi_1, \sigma_1}, \delta_{\xi_2, \sigma_2}] = \delta_{[\xi_1, \xi_2],\,\xi_1 \cdot \nabla_{\gamma_0} \sigma_2 - \xi_2 \cdot \nabla_{\gamma_0} \sigma_1}$. Together with the fact that $\Psi_{1\text{-loop}}[\gamma_0 + \sqrt{G}\,\mathfrak{h}]$ is invariant under the $\delta_{K_A, \,\sigma_A}$ and has no bosonic zero modes~\cite{Cotler:2025gui}, upon fixing the residual gauge transformations the one-loop norm is proportional to $1/\vol(SO(d,1))$, which equals zero.

It is natural to pursue the above argument to higher orders in $\sqrt{G}$.  For this, we need to account for the $O(\sqrt{G})$ terms on the right-hand sides of~\eqref{E:deltaF0} and~\eqref{E:deltaH1}.  With these new terms, the ghost zero modes will no longer be the ordinary CKVs, but rather correspond to $\xi$'s and $\sigma$'s which depend on $\mathfrak{h}$ and are thus \emph{field-dependent}.  A particular set of all-order, perturbative solutions to higher-loop equations was obtained in~\cite{Cotler:2025gui}, but that approach does not readily show if the resulting field-dependent $\text{diff} \times \text{Weyl}$ transformations furnish a representation of $\mathfrak{so}(d,1)$.  Here we will take a different approach which will make manifest the desired features of the higher-loop ghost zero modes.

Letting $g \in SO(d,1)$, the associated conformal diffeomorphism of $\gamma_0$ satisfies $\varphi_g^* \gamma_0 = e^{2 \omega_g} \gamma_0$ for a smooth function $\omega_g$ on $\mathbb{S}^{d-1}$.  We choose the $\varphi_g$ so that $\varphi_{g_1 g_2} = \varphi_{g_2}\circ \varphi_{g_1}$, and hence $\varphi_{g_1}^*\circ \varphi_{g_2}^* = \varphi_{g_1 g_2}^*$ on tensor fields.  Now take $\gamma \in \mathcal{S}$, and recalling that $\gamma = \gamma_0 + \sqrt{G}\,\mathfrak{h}$, we treat $\mathfrak{h} = O(G^0)$ and expand all quantities as power series in $\sqrt{G}$. Then $\varphi_g^*$ acts on $\gamma$ as
\begin{align}
\label{E:gammapush1}
\varphi_g^* \gamma = e^{2 \omega_g}(\gamma_0 + \sqrt{G}\,\mathfrak{h}_g)\,,\quad \mathfrak{h}_g \equiv e^{-2 \omega_g}\, \varphi_g^* \mathfrak{h}\,.
\end{align}
Here $\mathfrak{h}_g$ is traceless with respect to $\gamma_0$, but is not generally transverse.  As such, while $\gamma$ is on the gauge-fixing slice, $\varphi_g^* \gamma$ is generally not.  However, we will show that we can pick some $\xi_g[\gamma]$ and $\sigma_g[\gamma]$, each depending on $g$ and $\gamma$, such that the metric
\begin{align}
\label{E:Req1}
\mathcal{R}_g[\gamma] \equiv e^{2 \sigma_g[\gamma]} e^{\mathcal{L}_{\xi_g[\gamma]}} \,\varphi_g^* \gamma
\end{align}
\emph{is} in $\mathcal{S}$, to all orders in $\sqrt{G}$.  That is, $\xi_g[\gamma]$ and $\sigma_g[\gamma]$ are compensating diffeomorphisms and Weyl rescalings which bring $\varphi_g^* \gamma$ back to the gauge-fixing slice, but not to $\gamma$ (unless $g = \mathds{1}$).  We will further show that to all orders in $\sqrt{G}$, in a suitable prescription the $\xi_g[\gamma]$ and $\sigma_g[\gamma]$ can be chosen \emph{uniquely}, and hence $\mathcal{R}_g$ is a \emph{unique} map at all orders in $\sqrt{G}$ as well.

Before establishing uniqueness, let us examine what it will buy us.  Suppose that $\mathcal{R}_g$ is uniquely prescribed in the sense explained above, and that it satisfies $\mathcal{R}_g[\gamma_0] = \gamma_0$.  We first establish that $\mathcal{R}_g$ defines a nontrivial left action of $SO(d,1)$ on $\mathcal{S}$, and in particular $\mathcal{R}_{g_1} \circ \mathcal{R}_{g_2} = \mathcal{R}_{g_1 g_2}$.  By definition $\mathcal{R}_{g_1}(\gamma) = u_{g_1}[\gamma] \circ \varphi_{g_1}^* \gamma$ and $\mathcal{R}_{g_2}(\gamma) = u_{g_2}[\gamma] \circ \varphi_{g_2}^* \gamma$, where $u_{g_1}[\gamma]$ and $u_{g_2}[\gamma]$ are gauge transformations.  Composing $\mathcal{R}_{g_1}$ and $\mathcal{R}_{g_2}$, we have
\begin{align}
(\mathcal{R}_{g_1} \circ \mathcal{R}_{g_2})[\gamma] &= u_{g_1}[\mathcal{R}_{g_2}[\gamma]] \circ \varphi_{g_1}^*(\mathcal{R}_{g_2}[\gamma]) \\
&= u_{g_1}[\mathcal{R}_{g_2}[\gamma]] \circ \varphi_{g_1}^*(u_{g_2}[\gamma] \circ \varphi_{g_2}^*(\gamma))\,.
\end{align}
Inserting the identity $(\varphi_{g_1}^*)^{-1} \circ \varphi_{g_1}^* = \text{Id}$ and regrouping terms, we find
\begin{align}
\label{E:ucomp1}
u_{g_1}[\mathcal{R}_{g_2}[\gamma]] \circ (\varphi_{g_1}^* \circ u_{g_2}[\gamma] \circ (\varphi_{g_1}^*)^{-1}) \circ  (\varphi_{g_1}^* \varphi_{g_2}^*)(\gamma)\,.
\end{align}
Since $\varphi_{g_1}^* \circ u_{g_2}[\gamma] \circ (\varphi_{g_1}^*)^{-1}$ is a gauge transformation, the composite
\begin{align}
u_{12}[\gamma] \equiv u_{g_1}[\mathcal{R}_{g_2}[\gamma]] \circ \big(\varphi_{g_1}^* \circ u_{g_2}[\gamma] \circ (\varphi_{g_1}^*)^{-1}\big)
\end{align}
is also a gauge transformation.  Using $\varphi_{g_1}^* \varphi_{g_2}^* = \varphi_{g_1 g_2}^*$,~\eqref{E:ucomp1} becomes
\begin{align}
(\mathcal{R}_{g_1} \circ \mathcal{R}_{g_2})[\gamma] = u_{12}[\gamma] \circ \varphi_{g_1 g_2}^*(\gamma)\,.
\end{align}
By uniqueness of $\mathcal{R}_g$, we must have
\begin{align}
(\mathcal{R}_{g_1} \circ \mathcal{R}_{g_2})[\gamma] = \mathcal{R}_{g_1 g_2}[\gamma]\,,
\end{align}
so $\mathcal{R}_g$ defines a nontrivial left action of $SO(d,1)$ on $\mathcal{S}$.

The above in fact implies that the $\mathcal{R}_g$'s generate the manifold of ghost zero modes to all perturbative orders in $\sqrt{G}$; indeed the $\mathcal{R}_g$'s preserve the gauge-fixing conditions and form a nontrivial left action of $SO(d,1)$ on $\mathcal{S}$.  This is particularly visceral at the level of the corresponding Lie algebra: letting $g(t) = \exp(t\, K_A)$, we can define the $\text{diff} \times \text{Weyl}$ transformation $\delta_A$ by
\begin{align}
\delta_A \gamma \equiv \left.\frac{\dd}{\dd t}\right|_{t = 0} \mathcal{R}_{g(t)}[\gamma]\,,
\end{align}
and correspondingly
\begin{align}
\label{E:bettercomm1}
[\delta_A, \delta_B] = f_{\!AB}^{\,\,\,\,\,\,\,\,C} \,\delta_C\,,
\end{align}
which generalizes~\eqref{E:comm1} since the $\delta_A$'s are \emph{field-dependent}. As promised, these $\text{diff} \times \text{Weyl}$ transformations preserve $\mathcal{S}$ since
\begin{align}
0 = \left.\frac{\dd}{\dd t}\right|_{t = 0} F_i[\mathcal{R}_{g(t)}[\gamma]] = \delta_A F_i[\gamma]\quad\text{and}\quad 0 = \left.\frac{\dd}{\dd t}\right|_{t = 0} H[\mathcal{R}_{g(t)}[\gamma]] = \delta_A H[\gamma]\,.
\end{align}

It remains to show that the $\mathcal{R}_g$'s are unique to all orders in $\sqrt{G}$.  To this end, we write~\eqref{E:Req1} as
\begin{align}
\label{E:Req2}
\mathcal{R}_g[\gamma] &= e^{2 \sigma_g[\gamma]} e^{\mathcal{L}_{\xi_g[\gamma]}} e^{2 \omega_g}(\gamma_0 + \sqrt{G}\,\mathfrak{h}_g) \\
&= e^{2 \Sigma_g[\gamma]} e^{\mathcal{L}_{\xi_g[\gamma]}} (\gamma_0 + \sqrt{G}\,\mathfrak{h}_g)
\label{E:Req3}
\end{align}
where we have used
\begin{align}
\label{E:Sigmadef}
\Sigma_g[\gamma] \equiv \sigma_g[\gamma] + e^{\mathcal{L}_{\xi_g[\gamma]}} \,\omega_g\,.
\end{align}
We impose the conditions
\begin{align}
\label{E:RgNorm}
\mathcal{R}_g[\gamma_0]=\gamma_0\,,\quad \xi_g[\gamma]=O(\sqrt{G})\,,\quad \Sigma_g[\gamma]=O(\sqrt{G})\,,
\end{align}
which implies $\xi_g[\gamma_0]=0$ and $\Sigma_g[\gamma_0]=0$, and hence $\sigma_g[\gamma_0] = -\omega_g$ from~\eqref{E:Sigmadef}.  We expand the compensators as formal series in $\sqrt{G}$, namely
\begin{align}
\label{E:CompExp1}
\xi_{g\,i}[\gamma] = \sum_{p=1}^{\infty} G^{p/2}\,\xi^{(p)}_{g\,i}[\mathfrak{h}]\,,\quad
\Sigma_g[\gamma] = \sum_{p=1}^{\infty} G^{p/2}\,\Sigma^{(p)}_{g}[\mathfrak{h}]\,,
\end{align}
and similarly we expand the metric $\mathcal{R}_g[\gamma]$ around $\gamma_0$ as
\begin{align}
\label{E:RgExp1}
\mathcal{R}_g[\gamma]_{ij} = \gamma_{0ij} + \sum_{p=1}^{\infty} G^{p/2}\,\mathcal{R}^{(p)}_{g\,ij}[\mathfrak{h}]\,.
\end{align}
While in our perturbative regime $\mathfrak{h}$ and $\mathfrak{h}_g$ can also be written in a $\sqrt{G}$ expansion, we will not need to write this out explicitly.  The gauge-fixing conditions $\mathcal{R}_g[\gamma]\in\mathcal{S}$ are then equivalent to
\begin{align}
\label{E:RgGaugeConds}
\nabla_{\!\gamma_0}^j \mathcal{R}^{(p)}_{g\,ij}=0\,,\quad
\gamma_0^{ij} \mathcal{R}^{(p)}_{g\,ij}=0
\quad \text{for all }p\geq 1\,.
\end{align}

In order to solve for $\xi_g[\gamma]$ and $\Sigma_g[\gamma]$ such that~\eqref{E:RgGaugeConds} holds, we first observe that at order $p$ the unknown coefficients $\xi^{(p)}_{g\,i}$ and $\Sigma^{(p)}_g$ enter $\mathcal{R}^{(p)}_{g\,ij}$ \emph{linearly}.  Indeed, expanding the exponentials in~\eqref{E:Req3} and using~\eqref{E:CompExp1}, any term in $\mathcal{R}_g[\gamma]$ that contains $\xi^{(p)}$ or $\Sigma^{(p)}$ more than once is of order $G^{(p+1)/2}$ or higher.  Therefore, for each $p\geq 1$ we may write
\begin{align}
\label{E:Hstruct}
\mathcal{R}^{(p)}_{g\,ij} = \mathcal{M}^{(p)}_{g\,ij}\!\!\left[\mathfrak{h},\{\xi^{(q)}_{g}\}_{q\le p-1},\{\Sigma^{(q)}_{g}\}_{q\le p-1}\right] +\nabla_{\!\gamma_0\,i}\xi^{(p)}_{g\,j}+\nabla_{\!\gamma_0\,j}\xi^{(p)}_{g\,i} +2\,\Sigma^{(p)}_{g}\,\gamma_{0ij}\,,
\end{align}
where $\mathcal{M}^{(p)}_{g\,ij}$ is a symmetric tensor determined entirely by $\mathfrak{h}_g$ and lower-order data.  In particular, $\mathcal{M}^{(1)}_{g\,ij}=\mathfrak{h}_{g\,ij}$.

Using the form of $\mathcal{R}^{(p)}_{g\,ij}$ identified in~\eqref{E:Hstruct}, we can now solve for $\xi^{(p)}_{g\,i}$ and $\Sigma^{(p)}_g$ at fixed $p$. Taking the trace of~\eqref{E:Hstruct} and imposing~\eqref{E:RgGaugeConds} gives
\begin{align}
\gamma_0^{ij}\mathcal{M}^{(p)}_{g\,ij} + 2\,\nabla_{\!\gamma_0}^i\xi^{(p)}_{g\,i} +2(d-1)\Sigma^{(p)}_{g} = 0\,,
\end{align}
so that
\begin{align}
\label{E:SigmaSolvep}
\Sigma^{(p)}_{g} = -\frac{1}{d-1}\!\left(\nabla_{\!\gamma_0}^i\xi^{(p)}_{g\,i} + \frac{1}{2}\,\gamma_0^{ij}\mathcal{M}^{(p)}_{g\,ij}\right).
\end{align}
Thus we can solve for $\Sigma_g^{(p)}$ given $\xi^{(p)}_{g\,i}$ and $\mathcal{M}^{(p)}_{g\,ij}$ (which depend on lower order terms).

Solving for $\xi^{(p)}_{g\,i}$ as a function of lower order terms is more involved.  For this, it is convenient to define the traceless part of $\mathcal{M}^{(p)}_{g\,ij}$,
\begin{align}
\label{E:JpDef}
\mathcal{J}^{(p)}_{g\,ij} \equiv \mathcal{M}^{(p)}_{g\,ij} -\frac{1}{d-1}\,\gamma_{0ij}\,\gamma_0^{k\ell}\mathcal{M}^{(p)}_{g\,k\ell}\,,
\end{align}
so that substituting~\eqref{E:SigmaSolvep} into~\eqref{E:Hstruct} yields the manifestly traceless form
\begin{align}
\label{E:Htraceless}
\mathcal{R}^{(p)}_{g\,ij} = \mathcal{J}^{(p)}_{g\,ij} + (\textsf{C}\xi^{(p)}_{g})_{ij}\,,
\end{align}
where $\textsf{C}$ is the conformal Killing operator defined above. Taking the divergence of~\eqref{E:Htraceless} and imposing~\eqref{E:RgGaugeConds} gives $\nabla_{\!\gamma_0}^j\mathcal{J}^{(p)}_{g\,ij} +\nabla_{\!\gamma_0}^j(\textsf{C}\xi^{(p)}_{g})_{ij} = 0$, which we can write as
\begin{align}
\label{E:DpEq}
(\textsf{D}\xi^{(p)}_{g})_i = \nabla_{\!\gamma_0}^j\mathcal{J}^{(p)}_{g\,ij}\,.
\end{align}

As recalled above, $\textsf{D}$ is positive semidefinite and its zero modes are precisely the ordinary CKVs. Because $\textsf{D}$ has a nontrivial kernel, the existence of solutions to~\eqref{E:DpEq} is governed by the Fredholm alternative. Since $\textsf{D}$ is elliptic on the compact manifold $\mathbb{S}^{d-1}$, the equation $\textsf{D}\xi=Y$ is solvable if and only if $Y$ is orthogonal (with respect to the $\gamma_0$ inner product on 1-forms) to $\ker(\textsf{D}^\dagger)$.  Due to the self-adjointness of $\textsf{D}$, we have $\ker(\textsf{D}^\dagger) = \ker(\textsf{D})$ and therefore the solvability condition for~\eqref{E:DpEq} reduces to showing that $\nabla_{\!\gamma_0}^j\mathcal{J}^{(p)}_{g\,ij}$ is orthogonal to all CKVs $K_A$. Fortunately for us, this condition holds identically since integration by parts gives
\begin{align}
\int_{\mathbb{S}^{d-1}}\!\!\dd^{d-1}x \sqrt{\gamma_0}\,K_A^{\,i}\nabla_{\!\gamma_0}^j\mathcal{J}^{(p)}_{g\,ij} = -\!\int_{\mathbb{S}^{d-1}}\!\!\!\dd^{d-1}x \sqrt{\gamma_0}\,(\nabla_{\!\gamma_0}^jK_A^{\,i}) \mathcal{J}^{(p)}_{g\,ij} = -\frac{1}{2}\!\int_{\mathbb{S}^{d-1}}\!\!\!\dd^{d-1}x \sqrt{\gamma_0}\,(\textsf{C}\,K_A)^{ij} \mathcal{J}^{(p)}_{g\,ij} = 0
\end{align}
which vanishes identically, where we have used that $\mathcal{J}^{(p)}_{g\,ij}$ is symmetric and traceless.  Therefore
\begin{align}
\label{E:FredholmRg}
\int_{\mathbb{S}^{d-1}}\!\dd^{d-1}x \sqrt{\gamma_0}\,K_A^{\,i}\,\nabla_{\!\gamma_0}^j\mathcal{J}^{(p)}_{g\,ij}=0 \quad \text{for all CKVs }K_A\,,
\end{align}
which is exactly the solvability condition for~\eqref{E:DpEq}. Hence~\eqref{E:FredholmRg} is satisfied for every $p\geq 1$, and the Fredholm alternative guarantees that a solution $\xi^{(p)}_{g\,i}$ to~\eqref{E:DpEq} exists at each order.

Solutions to~\eqref{E:DpEq} are not unique because one may always add a homogeneous solution, namely a CKV.  We fix this freedom by imposing, for each $p\geq 1$,
\begin{align}
\label{E:xiOrthRg}
\int_{\mathbb{S}^{d-1}}\!\dd^{d-1}x\,\sqrt{\gamma_0}\,\gamma_0^{ij}\,\xi^{(p)}_{g\,i}\,K_{A\,j}=0 \quad \text{for all CKVs }K_A\,,
\end{align}
With this convention, $\textsf{D}$ is invertible on the orthogonal complement of its kernel, and therefore~\eqref{E:DpEq} admits a \emph{unique} solution $\xi^{(p)}_{g\,i}$ at each order $p$.  Then~\eqref{E:SigmaSolvep} uniquely determines $\Sigma^{(p)}_g$ at the same order.  This completes the inductive step.

Starting from $p=1$ and iterating, we obtain a unique formal power series solution~\eqref{E:CompExp1} for $\xi_g[\gamma]$ and $\Sigma_g[\gamma]$ satisfying~\eqref{E:RgGaugeConds}, together with the conditions~\eqref{E:RgNorm}.  Moreover, $\sigma_g[\gamma]$ is uniquely determined by~\eqref{E:Sigmadef}.  Therefore the map $\mathcal{R}_g:\mathcal{S}\to\mathcal{S}$ defined in~\eqref{E:Req3} is uniquely determined to all orders in $\sqrt{G}$ in a neighborhood of $\gamma_0$.

The remaining input needed to complete the all-loop norm argument is that the loop-corrected Hartle-Hawking wavefunction $\Psi$, restricted to the transverse-traceless gauge slice, is invariant under the field-dependent residual transformations $\mathcal{R}_g$, or perturbatively under $\delta_A$.  Rather than verifying this order by order in perturbation theory, it is useful to phrase the expected statement in a way that makes clear why $\mathcal{R}_g$-invariance is natural.

The basic point is that the late-time Hartle-Hawking wavefunction is fundamentally a functional on the quotient of boundary data by boundary $\text{diff}\times\text{Weyl}$ transformations.  This is the statement underlying the overlap~\eqref{E:IP1}, where the integration variable is really the conformal class $[\gamma]$ of the boundary metric.  A choice of gauge such as the transverse-traceless conditions $F_i=0$ and $H=0$ provides, perturbatively near $\gamma=\gamma_0$, a local section of this quotient, and the gauge-fixed wavefunction is simply the pullback of $\Psi$ along that section.  From this perspective, the ``residual'' transformations are not additional symmetries.  They simply reflect the fact that the round metric $\gamma_0$ has a nontrivial stabilizer in $\text{diff}\times\text{Weyl}$, namely $SO(d,1)$, which is generated by CKVs and their associated Weyl rescalings.  On the gauge-fixing slice this stabilizer is realized by the induced action obtained by composing a global conformal transformation with the compensating $\text{diff}\times\text{Weyl}$ transformation that returns to the slice; infinitesimally, this induced action is generated by the field-dependent transformations $\delta_A$ defined above.

Since the Hartle-Hawking wavefunction is defined by a bulk path integral with fixed boundary metric, it is invariant under boundary diffeomorphisms.  Classically, and perturbatively in a diffeomorphism-covariant renormalization scheme,\footnote{In $d \geq 4$ we implicitly treat gravity as an effective field theory, where UV divergences at higher loops are absorbed into an infinite tower of local higher-derivative (bulk and, where needed, boundary) counterterms. This does not affect the residual $SO(d,1)$ volume factor, which follows solely from $\text{diff}\times\text{Weyl}$ covariance of the renormalized construction.} the wavefunction depends on $\gamma$ only through its $\text{diff}\times\text{Weyl}$ equivalence class, up to the standard local boundary Weyl anomaly for odd bulk dimension.  In odd bulk dimensions, the Weyl anomaly does not obstruct invariance of the wavefunction under the finite-dimensional subgroup of conformal diffeomorphisms that fixes the conformal class of the round metric on the sphere.  Concretely, since acting on a general $\gamma \in \mathcal{S}$ by $\varphi_g^*\gamma$ need not preserve the transverse-traceless slice, we defined an induced action $\mathcal{R}_g$ on $\mathcal{S}$ by composing with the unique compensators $(\xi_g[\gamma],\sigma_g[\gamma])$ that restore $F_i = H = 0$.  Since $\mathcal{R}_g$ differs from $\varphi_g^*$ only by a $\text{diff}\times\text{Weyl}$ gauge transformation, we do not expect an anomaly or regulator effect that would spoil the induced $SO(d,1)$ invariance on the slice, and thus order by order in $\sqrt{G}$ we have $\Psi[\mathcal{R}_g[\gamma]] = \Psi[\gamma]$ for $\gamma \in \mathcal{S}$.

Finally, the quadratic fluctuation operator of $\Psi$ restricted to the transverse--traceless slice has no bosonic zero modes at $\mathcal{O}(G^0)$~\cite{Cotler:2025gui}.  Since on the compact slice the nonzero spectrum is discrete, a mode that is nonzero at leading order cannot become a \emph{continuous} bosonic zero mode at any finite order in the formal $\sqrt{G}$ expansion.  Perturbatively, the only continuous zero modes therefore occur in the ghost sector and are generated by the residual $SO(d,1)$ transformations induced by $\mathcal{R}_g$ on $\mathcal{S}$.  We now make explicit how quotienting by this residual redundancy produces the associated group volume factor $1/\vol(SO(d,1))$.

After imposing transverse-traceless gauge, the inner product reduces (perturbatively near $\gamma_0$) to an integral over $\mathcal{S}$ of the form
\begin{align}
\label{E:SliceInt_remark}
\int_{\mathcal{S}} [\dd\gamma]_{\mathcal{S}}\,\mathcal{I}[\gamma]\,,
\end{align}
where $[\dd\gamma]_{\mathcal{S}}\,\mathcal{I}[\gamma]$ denotes the full gauge-fixed measure-density on $\mathcal{S}$ (including the $\text{diff} \times \text{Weyl}$ Faddeev-Popov determinant and loop corrections).  This density is obtained by gauge-fixing a $\text{diff} \times \text{Weyl}$-invariant measure on the space of boundary metrics and restricting to the gauge slice.  Because $[\dd\gamma]_{\mathcal{S}}\,\mathcal I[\gamma]$ is the density induced by gauge-fixing a $\text{diff} \times \text{Weyl}$-invariant construction (with any required anomaly terms absorbed into $\mathcal{I}[\gamma]$) and because $\mathcal{R}_g[\gamma]$ remains in the same $\text{diff} \times \text{Weyl}$ orbit as $\gamma$, the induced density on $\mathcal{S}$ is $\mathcal{R}_g$-invariant in the sense of a pullback,
\begin{align}
\label{E:RgDensityInv}
(\mathcal{R}_g)^*\big([\dd\gamma]_{\mathcal{S}}\,\mathcal{I}[\gamma]\big) = [\dd\gamma]_{\mathcal{S}}\,\mathcal{I}[\gamma]\,,
\end{align}
at least in the regime of our formal $\sqrt{G}$-perturbation theory. One may then fix the residual $SO(d,1)$ redundancy by inserting the finite-dimensional Faddeev-Popov identity
\begin{align}
\label{E:resFP_remark}
1 = \Delta_{\rm res}[\gamma]\int_{SO(d,1)}\!\!\dd g
\prod_{A=1}^{\dim SO(d,1)}\!\! \delta\Big(\zeta_A(\mathcal{R}_g[\gamma])\Big),
\end{align}
where the $\zeta_A=0$ are any local conditions that intersect each $SO(d,1)$ orbit in $\mathcal{S}$ once, and $\Delta_{\rm res}[\gamma]$ is the corresponding determinant.  Using $\mathcal{R}_g$-invariance and changing variables $\gamma\mapsto\mathcal{R}_g[\gamma]$ in~\eqref{E:SliceInt_remark} yields the factorization
\begin{align}
\int_{\mathcal{S}}[\dd\gamma]_{\mathcal{S}}\,\mathcal{I}[\gamma] = \vol(SO(d,1))\int_{\mathcal{S} \,\cap\, \{\zeta_A =0\}}[\dd\gamma]_{\mathcal{S}}\,\mathcal{I}[\gamma]\,\Delta_{\rm res}[\gamma]\,,
\end{align}
where we have used $\int_{SO(d,1)}\!\dd g = \vol(SO(d,1))$. Thus, upon dividing by the residual group in the gauge-fixing procedure, the norm acquires an overall factor $1/\vol(SO(d,1))$ to all perturbative orders.

If we wanted to compute the coefficient multiplying $1/\vol(SO(d,1))$, among other things we would have to make explicit the finite Jacobian associated with the normalization of the residual parameters~\cite{Anninos:2020hfj} by introducing the $L^2(\gamma)$ Gram matrix of the vector fields $K_A[\gamma]$ in $\delta_{A} \equiv \delta_{K_A[\gamma],\,\sigma_A[\gamma]}$,
\begin{align}
G_{AB}(\gamma) \equiv \int_{\mathbb{S}^{d-1}}\!\dd^{d-1}x \sqrt{\gamma}\,\gamma_{ij}\,K_A^i[\gamma]\,K_B^j[\gamma]\,.
\end{align}
In perturbation theory we have $G_{AB}(\gamma) = G_{AB}(\gamma_0) + O(\sqrt{G})$ on the transverse-traceless slice.  The perturbative non-orthonormality only affects the finite Jacobian on the zero-mode subspace (equivalently, a finite rescaling absorbed into $\Delta_{\rm res}[\gamma]$), and does not modify the overall noncompact group-volume factor $\vol(SO(d,1))$.

\subsection{Coupling to matter}

We now explain why the preceding conclusions persist upon coupling gravity to matter, with the matter placed jointly in the no-boundary state.  Our arguments here will generalize those of~\cite{Cotler:2025gui} at one loop. For concreteness, let $\chi$ denote collectively all bulk matter fields and let $S[g,\chi]=S_{\rm EH}[g]+S_{\rm matt}[g,\chi]$ be the total action (with the usual boundary terms). The corresponding no-boundary wavefunction depends on the future boundary data for both the metric and the matter fields,
\begin{align}
\Psi[\gamma,\varphi] = \int_{\text{no}\,\text{bdy}}^{(\gamma,\varphi)} \frac{[\dd g]\,[\dd \chi]}{\text{diffs}}\,e^{iS[g,\chi]}\,,
\end{align}
where $\varphi$ denotes the induced boundary data for $\chi$ at conformal infinity.  The matter is required to be regular on the Euclidean cap (sometimes called the Bunch-Davies boundary condition, although this is often reserved for flat slicing), which selects for a joint gravity-matter state which is $SO(d,1)$-invariant.

The norm involves integrating over boundary data modulo the appropriate gauge redundancies.  For the present purposes the essential point is that boundary $\text{diff}\times\text{Weyl}$ acts on $(\gamma,\varphi)$ by the usual induced transformations on the metric and on the matter boundary data.  Since the gravitational gauge-fixing conditions we imposed in~\eqref{E:FHeqs1} depend only on $\gamma$, the Faddeev-Popov operator governing the $\text{diff}\times\text{Weyl}$ ghosts is unchanged by the presence of matter because the variation of the gauge-fixing functions with respect to $(\xi,\sigma)$ is identical to the pure-gravity case.  In particular, the perturbative ghost zero modes are still realized by the same field-dependent generators $\delta_A$.  Moreover, the same field-dependent generators act on the matter boundary data $\varphi$ by the standard $\text{diff}\times\text{Weyl}$ transformation laws.

The coupled gravitational-matter path integral is diffeomorphism covariant, and the smooth boundary condition on the Euclidean cap is de Sitter invariant. Thus $\Psi[\gamma,\varphi]$ depends on $(\gamma,\varphi)$ only through their $\text{diff}\times\text{Weyl}$ equivalence class (up to the familiar possibility of boundary Weyl anomalies in odd bulk dimension, which do not obstruct global $SO(d,1)$).  Pulling back to a gauge-fixing slice therefore yields a gauge-fixed wavefunction that is invariant under the induced residual transformations, acting on both the metric and the matter data by the same boundary $\text{diff}\times\text{Weyl}$ parameters $(\xi,\sigma)$. As in pure gravity, potential obstructions would require an anomaly for the global conformal group or a regulator that breaks diffeomorphism invariance in an irremovable way; these are absent in a diffeomorphism-covariant renormalization scheme for ordinary (non-diffeomorphism-anomalous) matter.

Then the appearance of $1/\vol(SO(d,1))$ follows entirely from (i) the existence of $SO(d,1)$ ghost zero modes associated with the stabilizer of the round boundary metric and (ii) invariance of the joint wavefunction under the corresponding residual transformations.  Matter fields do not remove this stabilizer; they only contribute additional (finite, after renormalization) determinant factors and interaction corrections.  Moreover, provided the coupled quadratic fluctuation problem about the Hartle-Hawking background does not introduce bosonic zero modes into the joint wavefunction,\footnote{We exclude cases with additional exact zero modes such as unfixed global shift symmetries or the well-known infrared subtleties of massless minimally coupled scalars.  These issues are logically independent of the residual $SO(d,1)$ volume factor that drives the vanishing of the norm.} the absence of a quadratic zero mode persists perturbatively order by order in $\sqrt{G}$.  Therefore the same gauge fixing of the residual $SO(d,1)$ produces the factor $1/\vol(SO(d,1))$ to all perturbative orders, and so the norm of the leading contribution to the gravity-matter no-boundary state vanishes.  This finding generalizes the one-loop result of~\cite{Cotler:2025gui}.

\section{Discussion}

We have succeeded in showing that the leading contribution to the no-boundary state has vanishing norm at all loops in perturbation theory, including when coupled to matter.  A more difficult question is addressing what exactly this means.

On the one hand, our results reinforce the finding of~\cite{Cotler:2025gui} that the norm of the leading contribution to the no-boundary state is not equal to the sphere partition function~\cite{Polchinski:1988ua, Anninos:2020hfj}.  One might have expected the results to agree since there is a na\"{i}ve picture of gluing the Hartle-Hawking geometry to its complex conjugate and having the Lorentzian parts cancel leaving a Euclidean sphere.  This is a `tree-level' intuition and indeed the norm and the sphere agree at tree level.\footnote{Relatedly, the norm of the Bunch-Davies state of a matter field on top of a fixed Hartle-Hawking background should agree with the matter sphere partition function on account of perturbative unitarity on the Lorentzian segment of the fixed background.}  However, the norm and the sphere do not agree at loop level, since among other differences the former is proportional to $1/\vol(SO(d,1))$ and the latter to $1/\vol(SO(d+1))$.  In hindsight the mismatch is not surprising since the integration contours of the two calculations differ, and moreover the bulk path integral which prepares the state involves an integration over complex metrics which do not cleanly separate into Lorentzian and Euclidean pieces.

Perhaps the most important question is whether our results have implications for cosmology, and the possibility of our living in the no-boundary state.  Indeed, the vanishing of the norm of the leading contribution to $\Psi$ suggests that the probability that our universe has sphere cross-section is zero.  Relatedly, our results show that the leading contribution to the no-boundary state has serious issues when it comes to perturbative cosmological correlators.  As emphasized in~\cite{Cotler:2025gui}, the zero norm means that cosmological correlators\footnote{Here the $O_i$ are understood as gauge-fixed late-time observables, i.e.~cosmological correlator insertions, in the same boundary gauge used to define the inner product.}
\begin{align}
\frac{\int \frac{[\dd\gamma]\,[\dd \varphi]}{\text{diff}\times\text{Weyl}} \Psi^*[\gamma, \varphi]\, O_1[\gamma, \varphi] \cdots O_k[\gamma, \varphi] \Psi[\gamma, \varphi]}{\int \frac{[\dd\gamma]\,[\dd \varphi]}{\text{diff}\times\text{Weyl}} |\Psi[\gamma, \varphi]|^2}
\end{align}
cannot be normalized in perturbation theory, as the numerator is expected to be finite with enough operator insertions~\cite{Chakraborty:2023yed, Chakraborty:2023los} (see also~\cite{Marolf:2008hg, Marolf:2012kh}).  It does not help to `normalize' the inner product by a $\vol(SO(d,1))$ as this will cancel out between the numerator and denominator.\footnote{One might still be tempted to `normalize' the inner product by a $\vol(SO(d,1))$ to at least render $\int \frac{[\dd\gamma]\,[\dd\varphi]}{\text{diff}\times\text{Weyl}} |\Psi[\gamma, \varphi]|^2$ as finite.  While we are in principle allowed to do so, we can only fix the normalization of the path integral once and for all; as such, this choice of $\vol(SO(d,1))$ would also multiply the norms of the subleading topologies of the no-boundary state (e.g.~$\mathbb{S}^1 \times \mathbb{S}^{d-2}$ or $\mathbb{T}^{d-1}$ cross-section).  These subleading topologies break $SO(d,1)$ down to compact groups, and so a `by hand' multiplicative factor of $\vol(SO(d,1))$ in the path integral will cause their norms to diverge.}  It is possible that non-perturbative gauge-fixing effects (e.g.~Gribov ambiguities) could impact our treatment, but it is not clear what these effects might be or if they would lead to larger issues that impact on standard gauge fixings in quantum cosmology.

It is particularly interesting to study the no-boundary state in the presence of e.g.~a slow-roll inflation (see e.g.~\cite{Maldacena:2024uhs, Chen:2024rpx} for recent discussions).  This joint state is also $SO(d,1)$-invariant and has no bosonic zero modes, and thus the norm has a $1/\vol(SO(d,1))$ factor; however, the inflaton wavefunction itself is non-normalizable~\cite{Cotler:2025gui}.  In any case, it is probably more natural to cut off the state before future asymptotia at the time when inflation ends, and then `glue' on to a reheating region.  Doing so would helpfully break the boosts of the $SO(d,1)$ symmetry and thus would seem to render the would-be norm finite, but it is not presently known how to do such a calculation in a principled way at the quantum level.  Another approach would be to include an observer with large entropy that would likewise break $SO(d,1)$ boosts\footnote{More precisely, as explained in~\cite{Cotler:2025gui}, the model of an observer in~\cite{Witten:2023xze} with a continuous spectrum breaks $SO(d,1)$ down to $SO(d-1) \times SO(1,1)$, and a bosonic zero mode corresponding to a global time shift along the observer's worldline soaks up the $\vol(SO(1,1))$.}~\cite{Cotler:2025gui}, but this appears to be ad hoc and is not a part of the standard treatment of cosmological correlators.  A different direction would be to take seriously the possibility that the would-be `subleading' contribution to the no-boundary state with a different topology is in fact leading, and thus is deserving of more careful study; see~\cite{Turiaci:2025xwi, Ivo:2025yek} for some recent work.

What is clear is that there is much yet to be understood about quantum cosmology.  A takeaway is that the path integral is a reliable guide for subtleties about quantum cosmology which may be less apparent in other formalisms.  This takeaway is not exactly new, and was well-appreciated by practitioners of Euclidean quantum gravity 40 years ago~\cite{Gibbons:1994cg}.  However, in the intervening years, the development of string theory and holography have given us more confidence and prowess with the path integral in a wide variety of examples, so that now we may return to quantum cosmology with a new set of eyes.

\subsection*{Acknowledgements}

We thank Kristan Jensen, Juan Maldacena, and Edward Witten for valuable discussions, and Kristan Jensen and Juan Maldacena for comments on the manuscript. The author is supported by an Alfred P.~Sloan Fellowship.

\appendix

\section{Sphere partition function}
\label{App:sphere}

In the main text, we showed that the norm of the leading contribution to the no-boundary state is proportional to $1/\vol(SO(d,1))$ to all orders in perturbation theory.  In this Appendix we use the same methods to establish related statements for sphere partition functions in gravity and gauge theory.

\subsection{Gravity}
\label{App:sphere_gravity}

Here we explain why the gravity partition function on the round sphere $\mathbb{S}^d$~\cite{Polchinski:1988ua, Anninos:2020hfj} for $d \geq 3$ is proportional to $1/\vol(SO(d+1))$ to all orders in perturbation theory by taking the approach of Section~\ref{subsec:TT}.  Consider
\begin{align}
Z_{\mathbb{S}^d}=\int \frac{[\dd g]}{\text{diffs}}\,e^{-S_{\rm EH}[g]}\,,
\end{align}
expanded around the round saddle $g_{\mu\nu}=g_{0\mu\nu}+\sqrt{G}\,h_{\mu\nu}$. We fix diffeomorphisms using the de Donder gauge.  Defining a `bar' notation by
\begin{align}
\overline{h}_{\mu\nu}\equiv h_{\mu\nu}-\frac12\,g_{0\mu\nu} (g_0^{\rho\sigma}h_{\rho\sigma})\,,
\end{align}
we impose
\begin{align}
\label{E:AppDeDonder_new}
F_\mu[g]\equiv \nabla_{\!g_0}^{\nu}\overline{h}_{\mu\nu}=0\,,
\end{align}
which defines the de Donder gauge slice
\begin{align}
\mathcal{S}_{\text{dD}}\equiv \{g:\,F_\mu[g]=0\}\,.
\end{align}

At one loop, the ghost operator is the linearization of $F_\mu$ along the diffeomorphism orbit around $h = 0$.  Under an infinitesimal diffeomorphism $\delta_\xi g_{\mu\nu}=\mathcal{L}_\xi g_{\mu\nu}$, one finds at $h=0$
\begin{align}
\label{E:AppDeltaF_new}
\delta_\xi F_\mu\big|_{h=0} = -(\textsf{D}_{\text{dD}}\xi)_\mu\,, \quad (\textsf{D}_{\text{dD}}\xi)_\mu\equiv -\big(\nabla_{\!g_0}^2\xi_\mu+(d-1)\xi_\mu\big)\,,
\end{align}
where the operator $\textsf{D}_{\text{dD}}$ is elliptic and self-adjoint with respect to the $L^2(g_0)$ inner product on vector fields.  Its kernel is precisely the space of Killing vectors of $g_0$, which generate $\mathfrak{so}(d+1)$; we denote a basis by $K_A$.

The all-loop persistence of the residual $SO(d+1)$ is particularly transparent here because the stabilizer of $g_0$ inside $\text{diff}(\mathbb{S}^d)$ consists of genuine isometries (i.e.~Killing vectors as opposed to \emph{conformal} Killing vectors).  Let $r\in SO(d+1)$ and let $\varphi_r:\mathbb{S}^d\to \mathbb{S}^d$ be the corresponding isometry of $\mathbb{S}^d$ with metric $g_0$, so that $\varphi_r^* g_0 = g_0$.  Define the induced action on metrics by
\begin{align}
\label{E:AppRg_def_new}
\mathcal R_r[g]\equiv \varphi_r^* g\,.
\end{align}
Since $\varphi_r$ is an isometry of the background, the gauge-fixing functional~\eqref{E:AppDeDonder_new} is equivariant: for $g=g_0+\sqrt{G}\,h$ we have $\varphi_r^* g=g_0+\sqrt{G}\,(\varphi_r^* h)$, and moreover $\overline{\varphi_r^* h} =\varphi_r^*\overline{h}$ and $\nabla_{\!g_0}\big(\varphi_r^* T\big) = \varphi_r^*(\nabla_{\!g_0}T)$ for any tensor $T$, since $\varphi_r^* g_0=g_0$.  Consequently
\begin{align}
F_\mu[\mathcal R_r[g]] = \nabla_{\!g_0}^{\nu}\big(\overline{\varphi_r^* h}\big)_{\mu\nu} =\nabla_{\!g_0}^{\nu}(\varphi_r^*\overline{h})_{\mu\nu} =\big(\varphi_r^*\nabla_{\!g_0}^{\nu}\overline{h}_{\mu\nu}\big) =\varphi_r^*F_\mu[g]\,.
\end{align}
Thus if $F_\mu[g] = 0$ then $F_\mu[\mathcal{R}_r[g]] = 0$, meaning that if $g \in \mathcal{S}_{\text{dD}}$ then $\mathcal R_r[g] \in \mathcal{S}_{\text{dD}}$.  As such, $\mathcal R_r$ defines a genuine (and field-independent) action of $SO(d+1)$ on the gauge slice, with
\begin{align}
\mathcal R_{r_1}\circ \mathcal R_{r_2}=\mathcal R_{r_1r_2}\,,\quad \mathcal R_r[g_0]=g_0\,.
\end{align}
Infinitesimally, for $r(t) = \exp(t K_A)$ we obtain the residual transformations
\begin{align}
\delta_A g\equiv \left.\frac{\dd}{\dd t}\right|_{t=0}\mathcal R_{r(t)}[g]=\mathcal{L}_{K_A}g\,,
\end{align}
which satisfy the $\mathfrak{so}(d+1)$ commutation relations
\begin{align}
[\delta_A,\delta_B]=c_{\!AB}^{\,\,\,\,\,\,\,\,C}\,\delta_C\,,
\end{align}
with $c_{\!AB}^{\,\,\,\,\,\,\,\,C}$ the $\mathfrak{so}(d+1)$ structure constants.

Because the Einstein-Hilbert path integral is diffeomorphism invariant, the gauge-fixed integrand is the pullback of a diffeomorphism-invariant functional to the slice $\mathcal{S}_{\text{dD}}$, and hence is invariant under the induced residual action $\mathcal R_r$.  Moreover, after quotienting by diffeomorphisms the round sphere saddle has no continuous bosonic zero modes.  Since the remaining quadratic spectrum on the compact manifold is discrete and has no accumulation at zero, an eigenvalue that is nonzero at leading order cannot become an exact zero order by order in $\sqrt{G}$.  Therefore, by nearly identical arguments as those at the end of Section~\ref{subsec:TT}, fixing the residual $SO(d+1)$ yields an overall factor $1/\vol(SO(d+1))$ and this factor persists to all orders in perturbation theory.

\subsection{Yang-Mills}
\label{App:sphere_YM}

We now turn to Yang-Mills theory on the round sphere $\mathbb{S}^d$ with compact gauge group $G$, working on the trivial bundle and expanding around the trivial connection.  The Euclidean path integral is
\begin{align}
Z_{\mathbb{S}^d}^{\text{YM}}=\int \frac{[\dd A]}{\text{gauge}}\,e^{-S_{\text{YM}}[A]}\,, \quad S_{\text{YM}}[A]=\frac{1}{4 g_{\text{YM}}^2}\int_{\mathbb{S}^d}\dd^d x \sqrt{g_0}\,\tr(F_{ij}F^{ij})\,,
\end{align}
with $F=\dd A+A\wedge A$ and $g_0$ the round metric.  We rescale $A_i=g_{\text{YM}}\,a_i$ so that the kinetic term for $a$ is canonically normalized.  Infinitesimal gauge transformations with a $\mathfrak{g}$-valued parameter $\alpha$ act by
\begin{align}
\delta_\alpha a_i = \nabla_{\!g_0\,i}\alpha + g_{\text{YM}}[\alpha, a_i]\,,
\end{align}
where $\nabla_{\!g_0}$ acts only on the spacetime index.

We impose the Lorenz gauge condition
\begin{align}
\label{E:AppYMGauge_new}
\mathcal{C}[a]\equiv - \nabla_{\!g_0}^i a_i=0\,,
\end{align}
and varying $\mathcal{C}$ under $\delta_\alpha a_i$ gives
\begin{align}
\label{E:AppYMDeltaC_new}
\delta_\alpha \mathcal{C}[a] &= -\nabla_{\!g_0}^i\delta_\alpha a_i =-\nabla_{\!g_0}^2\alpha - g_{\text{YM}}\,\nabla_{\!g_0}^i[\alpha, a_i]\,.
\end{align}
Thus, at one loop around $a=0$, the Faddeev-Popov operator is $\textsf{D}_{\text{YM}}=-\nabla_{\!g_0}^2$ acting on adjoint-valued scalars.  This operator is elliptic and self-adjoint with respect to the $L^2(g_0)$ pairing
\begin{align}
\langle \alpha,\beta\rangle\equiv \int_{\mathbb{S}^d}\dd^d x \sqrt{g_0}\,\tr(\alpha\beta)\,,
\end{align}
and its kernel consists of constant $\mathfrak{g}$-valued functions.  Therefore the one-loop ghost zero modes are precisely the constant gauge parameters, namely the Lie algebra of the global group $G$, and fixing this residual redundancy produces an overall factor $1/\vol(G)$ in the Yang-Mills $\mathbb{S}^d$ partition function.

As before, the $1/\vol(G)$ persists to all loops by a similar mechanism as the gravity setting.  In particular, the background connection $a=0$ has a nontrivial stabilizer inside the gauge group, namely constant gauge transformations, and the Lorenz gauge condition~\eqref{E:AppYMGauge_new} is equivariant under this stabilizer.  Concretely, for a constant group element $g \in G$ we define the induced action on gauge fields by
\begin{align}
\label{E:AppYM_Rg_def}
\mathcal{R}_g[a]_i \equiv g\,a_i\,g^{-1}\,.
\end{align}
Since $g$ is constant on $\mathbb{S}^d$, we have
\begin{align}
\mathcal{C}[\mathcal{R}_g[a]]
=-\nabla_{\!g_0}^i(g a_i g^{-1})
=g\,(-\nabla_{\!g_0}^i a_i)\,g^{-1}
=g\,\mathcal{C}[a]\,g^{-1}\,,
\end{align}
and hence if $\mathcal{C}[a]=0$ then $\mathcal{C}[\mathcal{R}_g[a]]=0$, i.e.~$\mathcal{R}_g$ maps the Lorenz slice to itself.  The maps $\mathcal{R}_g$ form a (field-independent) representation of $G$ on the slice, with
\begin{align}
\mathcal{R}_{g_1}\circ \mathcal{R}_{g_2}=\mathcal{R}_{g_1 g_2}\,,\quad \mathcal{R}_g[0]=0\,.
\end{align}
Infinitesimally, writing $g(t)=\exp(tT_A)$ for a basis $T_A$ of $\mathfrak{g}$, we obtain residual transformations
\begin{align}
\delta_A a_i \equiv \left.\frac{\dd}{\dd t}\right|_{t=0}\mathcal{R}_{g(t)}[a]_i = [T_A,a_i]\,,
\end{align}
which preserve the gauge condition since $\delta_A \mathcal{C}[a]=[T_A,\mathcal{C}[a]]$ and thus vanish on $\mathcal{C}[a]=0$.  Equivalently, the constant gauge parameters remain exact ghost zero modes on the Lorenz slice to all perturbative orders, because setting $g_{\text{YM}}\alpha=T_A$ in~\eqref{E:AppYMDeltaC_new} gives $\delta_{T_A}\mathcal{C}[a]= [T_A, \mathcal{C}[a]]=0$ on~\eqref{E:AppYMGauge_new}.

Since the gauge-fixed functional depends only on the gauge orbit, its restriction to the slice~\eqref{E:AppYMGauge_new} is invariant under the induced residual global action $\mathcal{R}_g$.  Moreover, around the trivial connection on $\mathbb{S}^d$ there are no bosonic zero modes after quotienting by gauge transformations.  Indeed, for $d \geq 2$ one has $H^1(\mathbb{S}^d)=0$ so there are no nontrivial flat deformations, and the gauge-fixed quadratic operator on transverse one-forms has no zero modes.  As in the gravity discussion, the remaining quadratic spectrum on the compact manifold is discrete with no accumulation at zero, and thus an eigenvalue that is nonzero at leading order cannot become an exact zero order by order in perturbation theory.  Therefore fixing the residual global $G$ produces an overall factor $1/\vol(G)$ (see e.g.~the arguments at the end of Section~\ref{subsec:TT}), and this factor persists to all perturbative orders in $g_{\text{YM}}$.

We remark that the same considerations also apply to Chern-Simons theory~\cite{Witten:1988hf} on $\mathbb{S}^3$ in Lorenz gauge.  The leading Faddeev-Popov operator is again $-\nabla_{\!g_0}^2$ on adjoint scalars, so the residual gauge redundancy is the global group $G$, producing a factor $1/\vol(G)$ in perturbation theory.  For non-compact $G$, we have $\vol(G)=\infty$ and so the perturbative $\mathbb{S}^3$ partition function vanishes in this sense.  On three-manifolds $M$ with nontrivial flat moduli (e.g.~when $b_1(M)>0$ or for saddles with $H^1(M,\text{ad}A)\neq 0$), one must additionally treat bosonic zero modes associated with deformations of flat connections, and the stabilizer of a given saddle is generally a proper subgroup of $G$ (typically the center for irreducible flat connections), so the residual group-volume factor depends on the saddle.  On $\mathbb{S}^3$ we have $\pi_1(\mathbb{S}^3)=0$, and for the trivial connection $H^1(\mathbb{S}^3, \text{ad}A) = 0$, so the trivial flat connection is isolated and no additional zero modes arise.

\bibliography{refs}
\bibliographystyle{JHEP}

\end{document}